\newcommand{\CLASSINPUTtoptextmargin}{50pt}
\newcommand{\CLASSINPUTbottomtextmargin}{50pt}
\newcommand{\CLASSINPUTinnersidemargin}{41pt}
\newcommand{\CLASSINPUToutersidemargin}{41pt}
\documentclass[conference,9pt]{IEEEtran}
\IEEEoverridecommandlockouts

\usepackage[utf8]{inputenc} %
\usepackage[T1]{fontenc}    %
\usepackage[table,xcdraw]{xcolor}
\usepackage{pifont}
\usepackage[colorlinks,
    linkcolor=red,citecolor=citecolor,urlcolor=blue]{hyperref}       %

\usepackage{algorithm}
\usepackage{graphicx}
\usepackage{xcolor}
\usepackage{todonotes}
\usepackage{amssymb}
\usepackage{threeparttable}
\usepackage{multirow}
\usepackage{amsmath}
\usepackage{bm}
\usepackage{float}
\usepackage{amsthm}
\usepackage{soul}
\usepackage[noend]{algpseudocode}
\usepackage{enumitem} %
\usepackage[subrefformat=parens,labelformat=parens]{subfig}

\usepackage{xspace}

\definecolor{citecolor}{RGB}{34,139,34}
\definecolor{mydarkblue}{rgb}{0,0.08,1}
\definecolor{mydarkgreen}{rgb}{0.02,0.6,0.02}
\definecolor{mydarkred}{rgb}{0.8,0.02,0.02}
\definecolor{mydarkorange}{rgb}{0.40,0.2,0.02}
\definecolor{mypurple}{RGB}{111,0,255}
\definecolor{myred}{rgb}{1.0,0.0,0.0}
\definecolor{mygold}{rgb}{0.75,0.6,0.12}
\definecolor{myblue}{rgb}{0,0.2,0.8}
\definecolor{mydarkgray}{rgb}{0.,0.2,0.2}

\definecolor{lightred}{RGB}{255,235,235}
\definecolor{lightgreen}{RGB}{235,255,235}
\definecolor{lightblue}{RGB}{235,235,255}
\definecolor{lightcyan}{RGB}{235,255,255}
\definecolor{lightmagenta}{RGB}{255,235,255}
\definecolor{lightyellow}{RGB}{255,255,235}

\definecolor{qxkcolor}{RGB}{215,235,255}
\definecolor{softmaxcolor}{RGB}{230,235,255}
\definecolor{probxvcolor}{RGB}{255,255,235}

\definecolor{topkcolor}{RGB}{255,235,235}
\definecolor{zecolor}{RGB}{255,255,235}
\definecolor{dynacolor}{RGB}{235,255,255}

\definecolor{reviewcolor}{RGB}{0,0,200}

\newcommand{\calD}{\mathcal{D}}
\newcommand{\calO}{\mathcal{O}}

\newcommand{\calP}{\mathcal{P}}

\DeclareMathOperator*{\argmax}{argmax}

\theoremstyle{plain}

\theoremstyle{definition}

\algdef{SE}[DOWHILE]{Do}{doWhile}{\algorithmicdo}[1]{\algorithmicwhile\ #1}%

\newcommand{\squishlist}{
 \begin{list}{$\bullet$}
  { \setlength{\itemsep}{0pt}
     \setlength{\parsep}{3pt}
     \setlength{\topsep}{3pt}
     \setlength{\partopsep}{0pt}
     \setlength{\leftmargin}{1.5em}
     \setlength{\labelwidth}{1em}
     \setlength{\labelsep}{0.5em} } }
     
\newcommand{\squishend}{
  \end{list}  }

\newcommand{\MS}{\texttt{MDR-HDONN}\xspace}

\graphicspath{{./figs/}}
\usepackage{geometry}
\AtBeginDocument{%
  }

\usepackage{pifont}
\usepackage[normalem]{ulem}
\usepackage{xspace}
\algdef{SE}[DOWHILE]{Do}{doWhile}{\algorithmicdo}[1]{\algorithmicwhile\ #1}%

\begin{document}
\makeatletter
\newcommand*\mytitlefontsize{\fontsize{23}{15.5}\selectfont}
\makeatother

\title{
Multi-Dimensional Reconfigurable, Physically Composable Hybrid Diffractive Optical Neural Network
}

\author
{
Ziang Yin,
Yu Yao,
Jeff Zhang,
Jiaqi Gu\\
Arizona State University\\
{\small \emph{jiaqigu@asu.edu}}
}

\maketitle
\begin{abstract}
\label{abstract}
Diffractive optical neural networks (DONNs), leveraging free-space light wave propagation for ultra-parallel, high-efficiency computing, have emerged as promising artificial intelligence (AI) accelerators.
However, their inherent lack of reconfigurability due to fixed optical structures post-fabrication hinders practical deployment in the face of dynamic AI workloads and evolving applications.
To overcome this challenge, we introduce, \emph{for the first time}, a multi-dimensional reconfigurable hybrid diffractive ONN system (\MS), a physically composable architecture that unlocks a new degree of freedom and unprecedented versatility in DONNs. 
By leveraging full-system learnability, \MS repurposes fixed fabricated optical hardware, achieving exponentially expanded functionality and superior task adaptability through the differentiable learning of system variables.
Furthermore, \MS adopts a hybrid optical/photonic design, combining the reconfigurability of integrated photonics with the ultra-parallelism of free-space diffractive systems.
Extensive evaluations demonstrate that \MS has 
digital-comparable accuracy on various task adaptations with 74$\times$ faster speed and 194$\times$ lower energy.
Compared to prior DONNs, \MS shows exponentially larger functional space with 5$\times$ faster training speed, paving the way for a new paradigm of versatile, composable, hybrid optical/photonic AI computing.
We will open-source our codes.
\end{abstract}

\vspace{-5pt}
\section{Introduction}
\label{sec:Introduction}
Emerging technologies promise efficiency and performance breakthroughs, and optical neural networks (ONNs) are at the forefront of this movement. 
They are poised to revolutionize next-generation AI computing by harnessing the inherent parallelism and speed of light.
Diffractive ONNs (DONNs)~\cite{NN_PhysRevLett19, NP_ICCP24_Wei, NP_OPTICA24_Peng, NN_CHEN2021, NN_Science17_Lin} represent a compelling realization of optical AI hardware, employing free-space optical signal modulation, diffraction, and interference to achieve ultra-parallel intelligent information processing. 
Different from integrated photonic tensor cores (PTCs)~\cite{NP_NATURE2017_Shen, NP_PIEEE2020_Cheng, NP_NaturePhotonics2021_Shastri, NP_SciRep2017_Tait, NP_Nature2021_Xu, NP_NatureComm2022_Zhu, NP_ICCAD24_Gu} based on programmable photonic integrated circuits (PICs) designed for matrix-vector multiplication, free-space DONNs eliminate the dimension limitations of 2-D silicon chips and operate directly on light waves in 3-D free space, offering unprecedented parallelism through a global-view spatial linear operation. 

However, DONNs face a significant challenge in their lack of reconfigurability after fabrication. 
The fixed manufactured phase masks limit their adaptability to new ML tasks, hindering their deployment in real-world rapidly changing AI workloads~\cite{NP_IJCAI23_Cunxi, NP_ICCAD22_Cunxi}.
While research into reconfigurable phase masks based on tunable metasurfaces shows promise, these approaches remain in the immature stages. 
They struggle with limitations such as low endurance and reliability, which make them unsuitable for adoption in the near term.

To circumvent this device-level restriction, we shift our focus to the system architecture level. 
Optical systems offer intrinsic adjustable variables—like light wavelength, phase mask spacing, and element size that impact system transmission but remain largely unexplored. 
These variables are often manually adjusted based on designer experience, raising the question: \textbf{can these built-in hardware variables be automatically optimized for better expressivity?}
However, for true task adaptation, adjusting these scalar variables alone lacks sufficient degrees of freedom. 
A fundamental system redesign is needed to enable the necessary reconfigurability. In this work, we address this challenge: \textbf{how to construct a versatile DONN system with exponentially many reconfigurable functionalities by re-purposing the same set of fabricated phase masks?}
Furthermore, considering the strengths and weaknesses of both diffractive ONNs and integrated PTCs, we explore \textbf{how to hybridize free-space optics and integrated photonics organically} to leverage both the ultra-parallel processing capability of DONNs and the superior reconfigurability of integrated PTCs for joint neural network acceleration.

To answer those questions, we investigate the intrinsic learnability of DONNs from the system level and propose a multi-dimensional reconfigurable hybrid DONN design, dubbed \MS, with a physically composable architecture that can flexibly switch the orientation and placement order of phase masks for exponentially larger functional space.
The main contributions are as follows,
\squishlist
    {\item We present \emph{the first} in-depth analysis of multi-dimensional learnability in DONNs, introducing a physically composable hybrid optical system \MS for extreme multi-functionality.}
    {\item \textbf{Auto-Learned Multi-Dimensional System Variables}: we enable differentiable learning of DONNs across multiple dimensions (wavelength, spacing, orientation, permutation), unlocking an exponentially larger functional space with enhanced expressivity.}
    {\item \textbf{Hybrid Optical/Photonic System}: Our design synergistically integrates free-space diffractive optics and reconfigurable integrated photonics, leveraging the strengths of both technologies for joint NN acceleration.}
    {\item \textbf{Superior Task Adaptability}: Extensive evaluations show that our system and differentiable learning algorithms enable digital-comparable task adaptation performance with 74 $\times$ faster speed and 194$\times$ lower energy, outperforming prior DONNs with exponentially higher functional space and 5$\times$ higher training efficiency.}
\squishend

\vspace{-10pt}
\section{Background}
\begin{figure}
    \centering
    \includegraphics[width=\columnwidth]{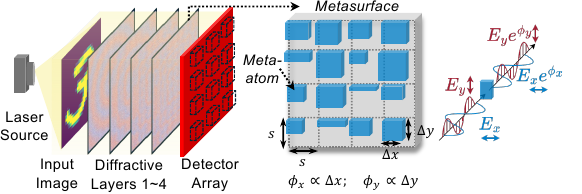}
    \caption{Diffractive optical neural networks (DONN) based on metasurfaces with a pixel size of $s$.
    The rectangular meta-atom shape can be designed to realize a polarization-dependent phase shift for orthogonally polarized light.
    }
    \vspace{-10pt}
    \label{fig:DONN}
\end{figure}

\label{sec:Background}
\subsection{Diffractive Optical Neural Networks (DONN)}

As shown in Fig.~\ref{fig:DONN}, the main trainable parameters are the modulation weights on the phase mask, i.e., $\bm{\Phi}\in\mathbb{R}^{K\times K}$, where the phase shift of one unit cell is $\bm{\phi_{i,j}}\in[0, 2\pi)$.
When incident light $X$ passes through the phase mask, it applies pixel-wise phase rotation, i.e., $e^{\Phi} \odot X$, where $\odot$ is element-wise multiplication.
Typically, to increase the number of trainable parameters, a DONN will have multiple cascaded phase masks, e.g., $L$ layers~\cite{NP_Science24_Lin}.
Between two phase masks, the coherent light will propagate in the free space with diffraction.
This can be described using the Rayleigh-Sommerfeld diffraction, a dense linear transformation where each light on the output plane is the superposition of all light from the input plane with a distance-related intensity/phase modulation coefficient $h$.
The transmission, including a phase mask and diffraction, is denoted as $X_{l+1}=\mathcal{H}(e^{\Phi_l}\odot X_l)$, formulated as~\cite{goodman2005introduction}:
\begin{equation}
    \label{eq:output_lightfield}
    \small
    \begin{aligned}
    &X_{l+1}(k_x^{l+1},k_y^{l+1})=\sum_{k_x}^K \sum_{k_y}^K h_{l}(R, z, \lambda)\cdot e^{\phi_{k_x^l,k_y^l}}\cdot X_l(k_x^l, k_y^l),\\
    &h(R, z, \lambda)= \frac{1}{2\pi} \frac{z}{R}(\frac{1}{R} - j\mathbf{k})\frac{e^{j\mathbf{k}R}}{R},\\
    &R = \sqrt{((k_x^{l+1} - k_x^{l})\cdot s)^2+((k_y^{l+1} - k_x^l) \cdot  s)^2+z^2},
    \end{aligned}
\end{equation}
In the diffraction coefficient $h$, $\mathbf{k} = \frac{2\pi}{\lambda} $ is the wave vector in free space, and $\lambda$ is the wavelength, and $z$ is the distance between two adjacent layers, $j = \sqrt{-1}$.
$R$ is the spatial distance from the unit cell located at $({k_x}^{l}, {k_y}^{l})$ in the $l$-th layer to the unit cell located at $({k_x}^{l+1}, {k_y}^{l+1})$ in the $(l + 1)$-th layer. 
$s$ is the pixel size of the phase mask unit cell.

As nanofabrication technology advances, a promising approach to implement DONN is using a sub-wavelength metasurface.
Metasurfaces offer a compact solution, where each unit cell is called a meta-atom and consists of sub-wavelength nanostructures~\cite{NP_ICCP24_Wei, NP_Nature2022_Luo, NP_Nature2024_Zheng}. 
A typical meta-atom is a rectangular nano-pillar with precisely defined width, length, and thickness to achieve a specific phase shift, in Fig.~\ref{fig:DONN}. 
This structure is polarization-dependent and thus can induce independent phase changes in orthogonal light polarizations, e.g., $\phi_x$ for x-polarized and $\phi_y$ for y-polarized light.
This is the foundation of our polarization-differential computing technique introduced later.

\subsection{Integrated Photonic Tensor Cores}
A different approach to implementing ONNs is through photonic tensor cores (PTCs). 
Unlike DONNs, where key components are passive diffractive components, PTCs integrate photonic devices to actively modulate light to achieve high-speed matrix-vector multiplication (MVM)~\cite{NP_NATURE2017_Shen, NP_Science2024_Xu, NP_HPCA2024_Zhu, NP_ACS2022_Feng,NP_NatureComm2022_Zhu,NP_SciRep2017_Tait, NP_Nature2021_Xu, NP_Nature2021_Feldmann}. 
DONNs are more natural for ultra-parallel low-energy spatial processing, while integrated PTCs are more suitable in generic reconfigurable MVMs.
This indicates complementary capabilities between two hardware platforms for joint computing.

\section{Related Works}
\label{sec:relatedwork}
Previous works have proposed solutions to build multi-functional DONNs.
For instance, prior methods have introduced rotation into diffractive neural networks, enabling certain layers to be rotatable and thereby enhancing task adaptability~\cite{NP_IJCAI23_Cunxi, NP_ICCAD22_Cunxi}. 
Prior work~\cite{NP_Nature2024_Zheng} also explored training different tasks on separate regions of the diffractive layer. 
This approach suffers from area limitations, constrained to a limited number of fixed functions (e.g., 2-3).
Similarly, metasurface-based DONNs~\cite{NP_Nature2022_Luo} leverage polarization-dependent phase responses to train multiple tasks, with each task corresponding to a different polarization state that the metasurface can distinguish. 
This approach faces limited expressivity as only a small number of orthogonal polarization states are supported by metasurfaces.

\section{Proposed Hybrid Diffractive Optical Neural Network \MS}
\label{sec:Method}
\begin{figure}
    \centering
    \includegraphics[width=0.99\columnwidth]{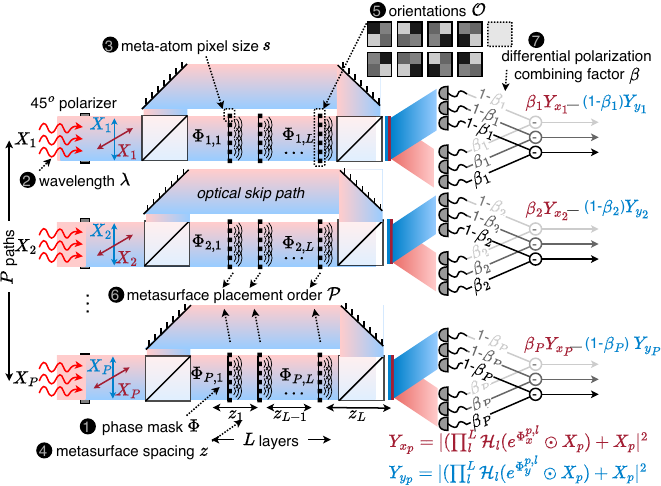}
    \vspace{-5pt}
    \caption{Multi-path diffractive layer (\texttt{DiffLayer}) with $P$ paths.
    Each has $L$ cascaded metasurfaces and polarization-differential photodetection.
    Learnable parameters \ding{202}-\ding{208} are marked.}
     \vspace{-5pt}
    \label{fig:MultipathDiffractive}
\end{figure}

We introduce a hybrid DONN with multi-dimensional reconfigurability and physical composability, dubbed \MS, shown in Fig.~\ref{fig:MultipathDiffractive}. 
Delving into the learnability of system-level parameters, our \MS brings new degrees of freedom to overcome the long-lasting reconfigurability limitations of traditional DONNs, enabling superior task adaptability and training efficiency.
\begin{table}[H]
\caption{Parameter tuning mechanism (structural, electrical, mechanical) and reconfig. frequency.
\ding{202}-\ding{208} belong to diffractive optical hardware.
\ding{209} is on integrated photonic chips.}
\label{tab:param_table}
\vspace{-5pt}
\resizebox{1\columnwidth}{!}{
\begin{tabular}{l|c|c|c|c}
\hline
\textbf{Learnable System Parameters}                    & \multicolumn{1}{l|}{\textbf{Stru.}} & \multicolumn{1}{l|}{\textbf{Elec.}} & \multicolumn{1}{l|}{\textbf{Mech.}} & \multicolumn{1}{l}{\textbf{Reconfig. Freq.}} \\ \hline
\ding{202}~Phase Response $\Phi$                             &    $\surd$                                      &                                          &                                               & \cellcolor[HTML]{F19C99}Fixed after fab.        \\
\ding{203}~Source Laser Wavelength $\lambda$                          &                                          & $\surd$          &  & \cellcolor[HTML]{CCFF99}Tunable per task                       \\
\ding{204}~Meta-atom Pixel Size $s$                        &  $\surd$                                        &                                          &      & \cellcolor[HTML]{F19C99}Fixed after fab.                                                  \\
\ding{205}~Metasurface Spacing $z$                        &                                          &                                          &   $\surd$                     & \cellcolor[HTML]{CCFF99}Tunable per task                                \\
\ding{206}~Orientation $\mathcal{O}$                &                                          &           & $\surd$      & \cellcolor[HTML]{CCFF99}Tunable per task                  \\
\ding{207}~Placement Order $\mathcal{P}$                      &                                          &           & $\surd$        & \cellcolor[HTML]{CCFF99}Tunable per task                \\ 
\ding{208}~Differential Polarization Factor $\beta$                            &           &   $\surd$                                       &       & \cellcolor[HTML]{9AFF99}Runtime tunable                  \\ \hline
\ding{209}~Channel-Mixing Factors $\alpha_{pre}$, $\alpha_{post}$ &           &  $\surd$                                        &     & \cellcolor[HTML]{9AFF99}Runtime tunable                    \\
 \hline
\end{tabular}
}
\vspace{-15pt}
\end{table}

\subsection{Overview of Learnable System Parameters}
For a generic expressive DONN, our \MS is assumed to have a \textbf{multi-path} architecture with $P$ parallel paths, each with $L$ cascaded phase masks, shown in Fig.~\ref{fig:MultipathDiffractive}.
We summarize programming mechanisms and properties of all learnable variables in our \MS architecture in Table~\ref{tab:param_table}, including the \emph{conventional} parameter phase masks $\Phi$, and our \textbf{7 newly introduced dimensionalities}: light source wavelength $\lambda$, meta-atom pixel size $s$, diffraction distance between metasurfaces $z$, metasurface orientation $\mathcal{O}$, metasurface placement ordering $\mathcal{P}$, pre- and post-channel-mixing factor $\alpha_{pre}$ and $\alpha_{post}$, and differential polarization combining factor $\beta$, visualized in Fig.~\ref{fig:MultipathDiffractive}.

\subsubsection{\ding{202}~Phase Masks $\Phi$}
Phases are the most important learnable parameters in DONN.
Adjusting the rectangular meta-atom width and length can independently change its polarization-dependent phase responses $\phi_x,\phi_y\in[0, 2\pi)$ for x-polarized and y-polarized light.
Once a phase mask is trained and fabricated, it is fixed and cannot be reconfigured for task adaptation.

\subsubsection{\ding{203}~Light Source Wavelength $\lambda$}
\label{subsec:lambda}
Tuning wavelengths can effectively change the optical system response.
Given a broadband meta-atom, its phase response $\phi$ is almost constant within a wide spectral range.
Changing wavelength, instead, mainly impacts the diffraction function $\mathcal{H}(\cdot | \lambda)$.
A reasonable range for $\lambda$ is from 400 nm visible spectrum to infra-red 1600 nm.
Given a preferred $\lambda$, e.g., 532 nm, we consider local fine-tuning with $\pm 20 nm$ linewidth given the tunability of the current laser with a 0.1 nm tuning resolution.

\subsubsection{\ding{204}~Meta-Atom Pixel Size $s$}
\label{subsec:pixelsize}
The diffraction operator $\mathcal{H}(\cdot)$ is a function of pixel size $s$, which is the period of the metasurface grid.
Considering all metasurfaces have a shared scalar pixel size $s\in\mathbb{R}$, e.g., 300 nm, we are allowed to freely determine this variable during the initial design stage for better performance.
The pixel size is lower-bounded by the meta-atom width required to realize 2$\pi$ phase response plus minimum manufacturable linewidth, e.g., $s>  s_{2\pi,\lambda_0}\cdot \lambda/\lambda_0+\Delta s$.
For example, given a reference $\lambda_0$=532 nm, the meta-atom width needs to be $s_{2\pi,\lambda_0}$=180 nm to realize 2$\pi$ phase shift, then we assume a linear scaling w.r.t. $\lambda$.
With a $\Delta s$=20 nm gap between pixels, the pixel period needs to honor the constraint $s>200$ nm.

\subsubsection{\ding{205}~Metasurface Spacing $z$}
The spacing between two phase masks determines the diffraction behavior of the light, i.e., $\mathcal{H}(\cdot|z)$
Intuitively, a very close distance passes the light directly to the next metasurface with very weak diffraction and cross-pixel interference.
Then, the system becomes almost a local linear operator with a nearly diagonal transfer matrix.
An overly large spacing introduces all-to-all pixel interaction. 
However, most light spreads to spaces outside the next metasurface plane, which causes significant light energy and information loss, and the next metasurface ends up receiving a dim and blurred pattern.
Hence, a carefully optimized diffraction spacing $z$ is critical to the receptive field and expressivity of the DONN.
We explore a shared learnable spacing $z$ for all layers and independently adjustable spacings $(z_1,z_2,\cdots, z_L)$ for different layers.

\subsubsection{\ding{206}~Metasurface Orientation $\mathcal{O}$}
\begin{figure}
    \centering
    \includegraphics[width=\columnwidth]{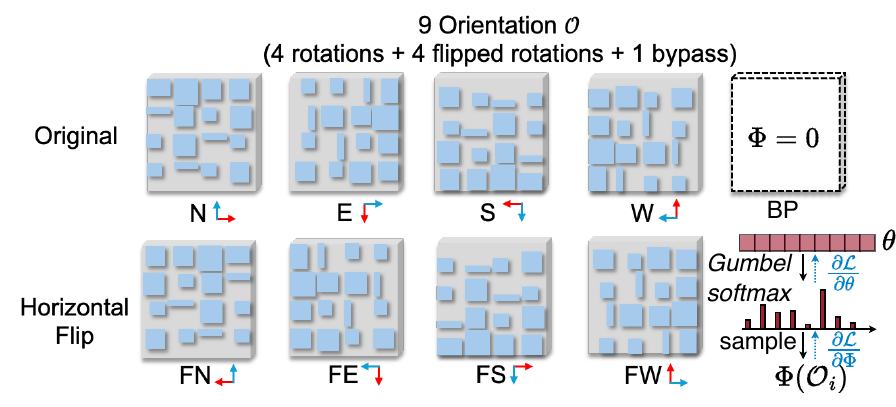}
    \vspace{-15pt}
    \caption{9 orientation states for each bi-directional phase mask, which can be learned differentiably using Gumbel-softmaxd.}
    \vspace{-5pt}
    \label{fig:Orientation}
\end{figure}

Before metasurface fabrication, the phase mask is freely optimized to realize arbitrary phase distribution.
After fabrication, the metasurface is fixed.
Thanks to the physical composability of our \MS design, the phase mask orientation becomes an additional degree of freedom to enrich the functionality while reusing the same piece of hardware, as shown in Fig.~\ref{fig:Orientation}.
The phase mask can be rotated in four angles, i.e., R0 (N), R90 (E), R180 (S), and R270 (W).
Besides, as a bidirectional phase mask, a metasurface can also be horizontally flipped to introduce chirality, i.e., shine the light reversely through the metasurface.
Then, we can augment to another 4 states: FN, FE, FS, and FW, shown in Fig.~\ref{fig:Orientation}.
Replacing the phase mask with an identity mask $\Phi=0$ is another allowable \emph{bypass} state given the system flexibility, denoted as BP.
Orientation reconfiguration can be implemented mechanically. 
Each metasurface can select its orientation out of 9 states, which gives \emph{exponentially many distinct hardware transmissions}.
For example, in a DONN with $P$ parallel paths, each with $L$ cascaded masks, there are $9^{PL}$ states.

Though this approach might not be suitable for real-time frequent reprogramming, it is promising for less frequent task adaptation. 
It enables \textbf{exponentially many new system responses for potential lifelong hardware reusing}.
Later, we will introduce how to learn the best state from discrete orientations in a differentiable way.

\subsubsection{\ding{207}~Metasurface Placement Order $\mathcal{P}$}
To fully leverage the physical composability of our \MS system, we further explore the location swapping of metasurfaces for extreme multi-functionality.
Inspired by the concept of card shuffling, phase masks can be reordered and re-plugged to any valid slots.
A permutation of all phase masks can be used to describe this placement ordering.
In a $P\times L$ multi-path DONN, the total permutation reaches $(PL)!$, enabling an even larger design space than orientation.

\subsubsection{\ding{208}~Differential Polarization Combining Factor $\beta$}
Polarization-multiplexed DONN~\cite{NP_Nature2022_Luo} is previously demonstrated to map two tasks onto orthogonal polarization channels.
Differently, we use two independent polarization directions in a \textbf{differential fashion to enhance the linear operation expressivity and realize full-range outputs}.
Each forward pass through our \MS system generates two non-negative outputs $Y_x$ and $Y_y$ for x(y)-polarized channel, respectively.
We use balanced photodetection to obtain a full-range feature map $Y=\beta Y_x-(1-\beta)Y_y$ with \emph{a non-negative combining factor} $\beta\in\mathbb{R}_+$. 
$\beta$ is electronically implemented with low overhead and thus can be dynamically reconfigured in real-time.

\subsubsection{\ding{209}~Pre/Post Channel-Mixing Coefficients $\alpha_{pre}$ and $\alpha_{post}$}
Simulating such systems in a multi-channel DONN layer is time-consuming and memory-hungry.
With $C_{in}$ input feature map channels and $C_{out}$ output channels, it requires to simulate the $P$-path \MS system by $C_{in}C_{out}/P$ times, which is not scalable.
To make DONN training more scalable and efficient, we borrow the concept of depth-wise convolution, which performs \textbf{spatial-only channel-wise diffractive projection} without aggregating the computing results from $C_{in}$ channels.
To enable channel-wise feature extraction, we add a channel-mixing coefficient $\alpha_{pre}$ and $\alpha_{post}$ before and after the metasurface DONN layer, same as point-wise convolution.
The pre-channel-mixing layer projects the features from $C_{in}$-channel to $C_{mid}$-channel.
The DONN system forwards $C_{mid}$ channels of images through the metasurface system and obtains $C_{mid}$ output channels, which are further fed into $\alpha_{post}$ layer and projects to $C_{out}$ channels.

As the channel-mixing operation only involves parallel MVM, it can be efficiently mapped to integrated PTCs for ultra-fast processing with real-time reconfigurability.
By leveraging the ultra-parallel global-view spatial processing capability of diffractive optical systems and ultra-fast reconfigurable MVM power of integrated photonic accelerators, our hybrid optical-photonic \MS system can realize \textbf{highly adaptable and efficient} optical computing.
Meanwhile, this depthwise separable operation, with comparable expressivity to traditional multi-channel convolution-like DONN layers, \textbf{significantly lowers the training cost} of simulating diffractive layers.

\subsubsection{Hybrid DONNLayer}
\begin{figure}[htp]
    \centering
    \vspace{-7pt}
    \includegraphics[width=\columnwidth]{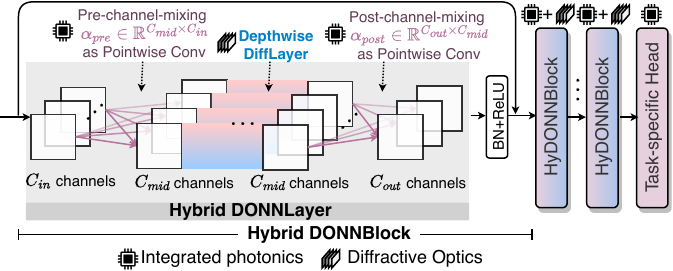}
    \vspace{-10pt}
    \caption{Hybrid DONNLayer with pointwise Convolution mapped to integrated PTCs and depthwise diffractive layers mapped to free-space diffractive optics.}
    \label{fig:DONNModel}
\end{figure}

Figure~\ref{fig:DONNModel} shows our hybrid \texttt{DONNLayer} with the channel-mixing operation mapped to integrated PTCs for ultra-fast point-wise Conv and diffractive layers (DiffLayer) in the middle mapped to metasurface systems for ultra-parallel depth-wise spatial information extraction.
\texttt{DONNResBlock} contains BatchNorm, activation, and a residual connection.
Multiple \texttt{DONNResBlocks} construct the backbone of our \MS system.
The task-specific head at the end is mapped on integrated PTCs.

\subsection{Pre-Fab Training for Initial Task}
In the initial training phase, conducted before metasurface fabrication, all system variables can be freely optimized.
Note that since the phase response of the metasurfaces $\Phi$ can be freely adjusted, it is unnecessary to learn the orientation or placement order at this stage. 
The permutation $\mathcal{P}$ and metasurface rotation mask $\mathcal{M}$ will not be optimized during this phase. 
Therefore, we focus on optimizing parameters such as the $\Phi$, $\alpha_{pre}$, $\alpha_{post}$, $\beta$, $\lambda$, $z$, and $s$ on the initial machine learning task. 
\begin{equation}
\small
\label{eq:InitialLearning}
\min_{\Phi,\lambda,z,s,\alpha,\beta}~\mathcal{L}((\Phi,\lambda,z,s,\alpha,\beta), \mathcal{D}_{init}).
\end{equation}

\subsection{Post-Fab Training for Tasks Adaptation}
Once the first training is completed, DONN phase masks will be fabricated.
Phases $\Phi$ and pixel size $s$ are fixed.
We now introduce how to enable fully-differentiable learning of the rest of the architecture variables to adapt the fabricated system to new tasks $\calD_{adapt}$,
\begin{equation}
\small
\label{eq:TaskAdaptaion}
\begin{aligned}
    \min_{\lambda,z,\alpha,\beta,\mathcal{O}, \mathcal{P}}&~\mathcal{L}((\lambda,z,\alpha,\beta,\mathcal{O}, \mathcal{P}), \mathcal{D}_{adapt}|\Phi^*,s^*).\\
    \text{s.t.}~~~&\mathcal{P}\in \text{Permutation matrices}\\
    &\lambda^*-\Delta \lambda \le \lambda \le \lambda^*+\Delta \lambda,~~\beta \ge 0.
\end{aligned}
\end{equation}

\subsubsection{Adapt Wavelength $\lambda$}
Similar to the initial training, we allow local adjustment of wavelength with $\pm 20$nm range to adapt the diffraction operation to new tasks.

\subsubsection{Adapt Metasurface Spacing $z$}
Similar to the initial training stage, we can either learn a shared spacing $z\in \mathbb{R}$ or dedicated layer-wise spacing vector $\mathbf{z}\in\mathbb{R}^L$ with standard gradient descent method.

\subsubsection{Learn Metasurface Orientation $\mathcal{O}$ with Gumbel-Softmax Method}
\label{sec:LearnOrientation}
Selecting the orientation of metasurfaces is a highly discrete problem.
We model the orientations of each metasurface as a 9-category parametric distribution.
For the $(p,l)$-th metasurface, its probability of having orientation $\mathcal{O}_{p,l}\in\mathbf{O}=\{N,W,S,E,FN,FW,FS,FE,BP\}$ is
\begin{equation}
    \small
    \label{eq:OrientationProb}
    P_{\theta_{p,l}}\big(\mathcal{O}_{p,l}=\mathbf{O}[i]\big)=e^{\theta_{p,l}^i}\Big/\sum_{i=1}^9e^{\theta_{p,l}^i}.
\end{equation}
We employ Gumbel-softmax to learn the distribution parameter $\theta$ with a gradient-based method in Fig.~\ref{fig:Orientation}.
The phase mask $\Phi$ is the superposition of all 9 states weighted by a mask $m$, i.e., $\Phi_{p,l}=\sum_{i=1}^9m_{p,l}^i \Phi_{p,l}^i$, where the mask is sampled from the following distribution,
\begin{equation}
    \small
    \label{eq:OrientationForward}
    \begin{aligned}
        m_{p,l}^i=\texttt{GumbelSoftmax}(\theta_{p,l}^i|\theta_{p,l})=\frac{e^{(\theta_{p,l}^i+g_{p,l}^i)/\tau}}{\sum_{i}e^{(\theta_{p,l}^i+g_{p,l}^i)/\tau}},
    \end{aligned}
\end{equation}
where $g$ and $\tau$ are the Gumbel noise and temperature.
The final orientation $\mathcal{O}_{p,l}$ can be selected by $\mathcal{O}_{p,l}=\mathbf{O}[\argmax_i \theta_{p,l}^i]$.

\subsubsection{Learn Metasurface Placement Order $\mathcal{P}$ with Differentiable Permutation Learning}
\label{sec:Permutation}
\begin{figure}
    \centering
    \vspace{-10pt}
    \includegraphics[width=0.9\columnwidth]{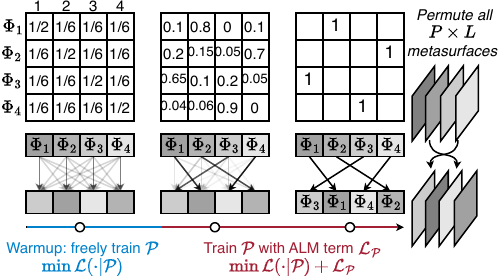}
    \vspace{-6pt}
    \caption{Illustration of metasurface placement order permutation. Two-stage training is adopted with warmup training and ALM-based permutation learning.}
    \vspace{-5pt}
    \label{fig:Permutation}
\end{figure}

We introduce an automatic learning method to permute all $PL$ metasurfaces in a differentiable way.
Without loss of generality, we use the learning of a permutation across $P$ metasurfaces as an example.
Exploring the factorial permutation space is very challenging.
Inspired by a differentiable circuit routing learning algorithm~\cite{NP_DAC2022_Gu}, as illustrated in Fig.~\ref{fig:Permutation}, we use an $P\times P$ permutation matrix $\mathcal{P}$ to represent the mapping from metasurface indices to slot indices.
A permutation matrix is a Boolean matrix $\mathcal{P}\in\{0,1\}^{P\times P}$ where each row and column only has one 1.
$P_{ij}=1$ means the $i$-th metasurface is placed at slot $j$.
We convert the permutation constraint with its equivalent continuous format, i.e., the row-wise and column-wise $\ell_1$-norm equals to squared $\ell_2$-norm.
We solve the constrained optimization problem with augmented Lagrangian method (ALM) to relax the constraint as an objective term,
\begin{equation}
\small
    \label{eq:sw-alm_term}
    \begin{aligned}
        \mathcal{L}_{\mathcal{P}} = \sum_{i=1}^{P} \mu_{i,r}^T \big(\Delta \widetilde{\mathcal{P}}^{i,:} + \frac{\rho}{2}(\Delta \widetilde{\mathcal{P}}^{i,:})^2\big) 
        + \mu_{i,c}^T \big(\Delta \widetilde{\mathcal{P}}^{:,j} + \frac{\rho}{2}(\Delta \widetilde{\mathcal{P}}^{:,j})^2\big),
    \end{aligned}
\end{equation}
where $\mu^r\in\mathbb{R}^P$ and $\mu^c\in\mathbb{R}^P$ are row-wise and column-wise Lagrangian multipliers, and $\rho$ is the quadratic penalty weight.
$\Delta \widetilde{\calP}^{i,:}=\|\widetilde{\calP}^{i,:}\|_1-\|\widetilde{\calP}^{i,:}\|_2^2$.
At every step when we update $\calP$, we also perform the dual update on the multiplier $\mu \gets \mu+\rho(\Delta \widetilde{\calP}^{i,:}+\frac{1}{2}(\Delta \widetilde{\calP}^{i,:})^2)$ to gradually increase the penalty to push it to a feasible permutation.

\subsubsection{Train Task-specific Head Layers}
After the backbone, task-specific head layers, such as the final linear classifier or convolutional regression layers, are re-initialized and trained on the new task.

\section{Evaluation Results}

\subsection{Settings}

\subsubsection{Model and Dataset}
As a case study, we assume our \MS system has up to 4 paths and up to 4 cascaded metasurface layers.
The DONN model backbone based on this hardware contains two Hybrid \texttt{DONNBlocks}. 
In all \texttt{DONNBlocks}, the depthwise \texttt{DiffLayers} are mapped to the same \MS hardware.
All structural and mechanical parameters that cannot be switched in runtime, i.e., $\lambda$, $z$, $s$, $\Phi$, $\calO$, $\calP$, \textbf{are shared} across all \texttt{DiffLayers}.
Runtime reconfigurable parameters, i.e., $\alpha_{pre}$, $\alpha_{post}$, and $\beta$, \textbf{are independently learned} for each \texttt{DiffLayer}.
The metasurface has 32$\times$32 meta-atoms. 
The initial values for wavelength $\lambda = 532nm$, meta-atom size $s = 400 nm$, and metasurface spacing $z = 8.42 \mu m$ are typical values used in the literature~\cite{NP_Nature2022_Luo}. 
We evaluate on three benchmarks: 
\begin{itemize}[leftmargin=*]
\setlength{\itemindent}{0.5em}
    \item Pre-fab training on Fashion-MNIST~\cite{NN_FashionMNIST2017} 10-class recognition and adapt the model to QuickDraw~\cite{NN_QuickDraw2017} 10-class image recognition. A classification head is used with AdaptiveAvgPool5-FC10.

    \item Pre-fab training on CIFAR-10~\cite{NN_cifar2009} and adapt the model to CIFAR-100~\cite{NN_cifar2009}. The classification head is Adaptive AvgPool5-FC10(100).

    \item Pre-fab training on Darcy flow~\cite{NN_DarcyFlow} 1-channel PDE solving and adapt to Navier-Stokes~\cite{NN_NaviorStoke} 10-channel PDE solving. A regression head is used with C64K3-BN-GELU-C1(10)K1.
\end{itemize}

\subsubsection{Training Settings}
\underline{General Setting}:
We train all DONNs for 100 epochs using the AdamW~\cite{NN_AdamW2017_Ilya} optimizer with a learning rate (lr) of 1E-2, a cosine decay scheduler, a weight decay rate of 1E-2, and data augmentation techniques such as random cropping and flipping for both pre-fabrication training tasks and adaptation tasks. 
Weight decay is only applied to $\Phi$ which is reparameterized in the range of $[-\pi, \pi)$. 

\noindent\underline{Schedule $\tau$ in Orientation and $\rho$ in Permutation}:~
For orientation learning, to prioritize early-stage exploration and later-stage stable convergence, we apply exponential decay on the Gumbel-softmax temperature $\tau = \tau_{init} \cdot \gamma_{\tau}^{(epoch - 1)}$, where $\tau_{init}$ is 5 and $\gamma_{\tau}$ is 0.956.
To encourage early-stage free permutation optimization and gradually force the constraints, ALM quadratic penalty term $\rho$ is scheduled as $\rho = \rho_{init} \cdot \gamma_{\rho}^{(epoch - warm\;up\;epoch)}$, with a warm-up period of 5 epochs, $\rho_{init}$ set to 1E-5, and $\gamma_{\rho}$ set to 1.2.

\noindent\underline{Parameter Quantization}:~
$\lambda$ is allowed to be adjusted within $\pm 20$nm around initial value, with 1 nm precision. 
To ensure that the meta-atom can generate a $-\pi$ to $\pi$ phase response for a given $\lambda$, the pixel size's lower bound is calculated following Sec.~\ref{subsec:pixelsize} equation: $s > \frac{\lambda }{\lambda_{0}} \cdot s_{2\pi, \lambda_0} + \Delta_s$, where $\lambda_{0} = 532$nm, $s_{2\pi, \lambda_0} = 300$nm, and $\Delta_s = 20$nm.
The precision for $z$ is set to 10 nm, while $\Phi$ is reparameterized $\in [-\pi, \pi)$ with a precision of $1^{\circ}$.
We also employ learned step-size quantization-aware training~\cite{NN_ICLR2020_Esser}. 
A $b_w$=8-bit quantization is applied to weights and activations in layers utilizing PTCs, such as $\alpha$ and the final head.

\noindent\underline{Evaluation Metrics}:~
Accuracy is used as the primary metric for the classification tasks. 
The Mean Squared Error (MSE) was utilized as the evaluation metric for PDE-solving tasks.
Both training and test performance are shown to indicate each setting and parameter's contribution to the DONN's expressivity, trainability, and generalizability.

\subsection{Ablation Study}
\label{sec:result}
\subsubsection{\MS Parameter Expressivity and Learnability in Pre-fab Training}
\begin{figure}
    \centering
    \includegraphics[width=\columnwidth]{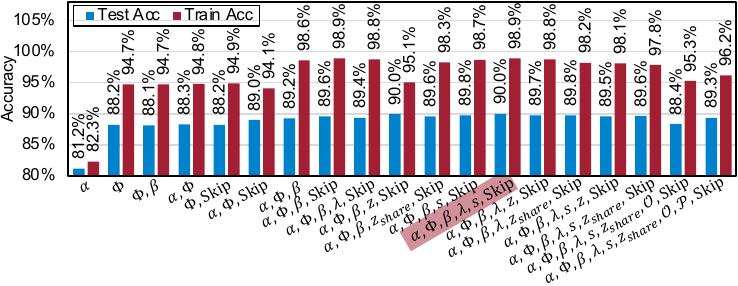}
    \vspace{-15pt}
    \caption{Expressivity of different combinations of learnable system parameters. Training and test accuracy are shown on 2-layer DONN on FMNIST.}
    \vspace{-10pt}
    \label{fig:ParamExpressivity}
\end{figure}

In initial training, we conduct a thorough ablation study on parameters in Fig.~\ref{fig:ParamExpressivity}.
We train ($\alpha$, $\Phi$, $\beta$, $\lambda$, $s$) with the optical skip path enabled as the optimal default pre-fab training settings.

\subsubsection{\MS \; \emph{Individual} Effectiveness of Params. \ding{203}-\ding{208} in Post-fab Task Adaptation}
\label{subsec: parameter effectiveness}
\begin{figure}
    \centering
    \vspace{-10pt}
    \subfloat[]{\includegraphics[width=0.49\columnwidth]{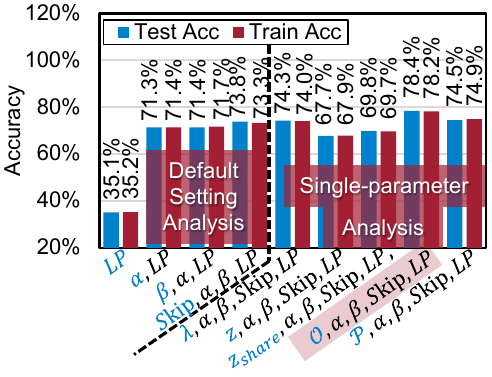}
    \label{fig:ParameterIsolation}}
    \subfloat[]{\includegraphics[width=0.512\columnwidth]{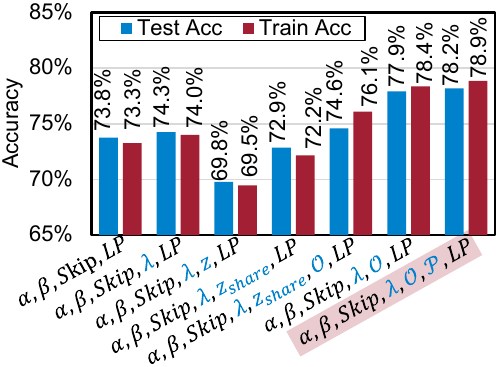}
    \label{fig:ProgressiveAdaptation}}
    \vspace{-6pt}
    \caption{
    (a) Effectiveness of individual parameter (\ding{203}-\ding{207}) for task adaptation, LP: Linear Probing. 
    (b) Effectiveness of progressively enabling parameters (\ding{203}-\ding{207}) in adaptation.}
    \vspace{-8pt}
\end{figure}

We explore the effectiveness of each parameter in task adaptation in Fig.~\ref{fig:ParameterIsolation}. 
Simply training the head, called linear probing (LP), with the backbone frozen is insufficient. 
As we enable \emph{electrically} tunable parameters ($\beta$ and $\alpha$), along with the optical skip path and linear probing, we observe a clear accuracy increase.

For \emph{mechanically} tunable parameters ($\lambda$, $z$, $\calO$, $\calP$) in Fig.~\ref{fig:ParameterIsolation}(\emph{Right}), orientation $\mathcal{O}$ provides additional $9^4$ dimensionality, resulting in a better performance for task adaptation. 
Permutation $\mathcal{P}$ also contributes positively to task adaptation, though not as impactful as orientation due to the $4!$ solution space, 273$\times$ smaller than $\mathcal{O}$. 
The effectiveness of $\mathcal{P}$ is expected to grow as the number of paths increases.

The diffractive distance $z$ and $z_{share}$
contribute less to performance, as they are too sensitive to be trained alongside other parameters, leading to an increase in training difficulty and thus impacting the overall optimality, which is not trained during adaptation.

\subsubsection{\MS \; \emph{Combinational} Effectiveness of Params. \ding{203}-\ding{208}  in Post-fab Task Adaptation}
\label{subsec:TaskAdaptationAbility}
After analyzing the effectiveness of proposed parameters individually, we progressively enable these parameters to evaluate their task adaptation ability, as shown in Fig.~\ref{fig:ProgressiveAdaptation}. 
The results are consistent with the result evaluated in Sec.~\ref{subsec: parameter effectiveness} that the diffractive distance $z$ is particularly sensitive and tends to reach a suboptimal state when trained alongside other parameters.
However, once $z$ is fixed to its initial learned value, we observe a clear accuracy improvement as purposed parameters, $\lambda$, $\mathcal{O}$, and $\mathcal{P}$, are progressively enabled during adaptation.

\subsubsection{\MS Orientation Effectiveness}
\label{subsec:Orientation}
\begin{figure}
    \centering
    \vspace{-5pt}
    \includegraphics[width=0.6\columnwidth]{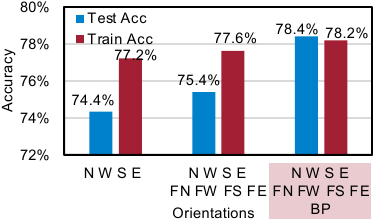}
    \vspace{-5pt}
    \caption{Nine orientations show the best expressivity and accuracy on QuickDraw adaptation tasks.}
    \vspace{-10pt}
    \label{fig:OrientationResults}
\end{figure}

Figure~\ref{fig:OrientationResults} provides strong evidence for the effectiveness of the 9-option orientations.
Compared to 4-direction rotation~\cite{NP_IJCAI23_Cunxi, NP_ICCAD22_Cunxi}, the extra 4 flipped states and the identity state shows +4\% higher test accuracy.

\begin{table}[t]
\centering
\caption{Evaluate \MS($P$=4, $L$=4) on pre-fab initial tasks and post-fab adapted tasks.
A 2-layer digital CNN (C32K3)$_{\times 2}$ is shown as a reference.
In each group, DONN is trained on the first task and adapted to the second task.
We use initial training results of the second task as a reference.
For classification, test Acc. is shown.
For Darcy/NavierStokes, MSE is shown.}
\label{tab:mainresults}
\vspace{-5pt}
\resizebox{1\columnwidth}{!}{
\begin{tabular}{l|c|c|c}
\hline
\multicolumn{1}{c|}{Tasks} & \begin{tabular}[c]{@{}c@{}}Digital 2-Layer CNN \\ C32K3-C32K3\end{tabular} & \begin{tabular}[c]{@{}c@{}}Pre-fab\\ Initial Training\\ $P$=4, $L$=4, $\alpha$, $\Phi$, \\ $\beta$, $Skip$, $s$, $\lambda$\end{tabular} & \begin{tabular}[c]{@{}c@{}}Post-fab\\ Task Adaptation\\ $P$=4, $L$=4, $\alpha$, $\beta$, $Skip$, \\$\lambda$, $\mathcal{O}$, $\mathcal{P}$\end{tabular} \\ \hline
FMNIST           & 0.9122                                                                       & \cellcolor[HTML]{E6E6E6}0.9014                                                                                        & \cellcolor[HTML]{E6E6E6}-                                                                                                               \\
Quickdraw        & 0.8978                                                                       & \cellcolor[HTML]{E6E6E6}0.8646                                                                                        & \cellcolor[HTML]{E6E6E6}0.7992 (adapt from FMNIST)                                                                                                         \\ \hline
CIFAR10          & 0.7633                                                                       & \cellcolor[HTML]{E6E6E6}0.6182                                                                                        & \cellcolor[HTML]{E6E6E6}-                                                                                                               \\
CIFAR100         & 0.4347                                                                       & \cellcolor[HTML]{E6E6E6}0.3815                                                                                        & \cellcolor[HTML]{E6E6E6}0.2513 (adapt from CIFAR10)                                                                                                         \\ \hline
Darcy           & 0.0466                                                                       & \cellcolor[HTML]{E6E6E6}0.055                                                                                         & \cellcolor[HTML]{E6E6E6}-                                                                                                               \\
NavierStokes    & 0.1085                                                                       & \cellcolor[HTML]{E6E6E6}0.1007                                                                                        & \cellcolor[HTML]{E6E6E6}0.1753 (adapt from Darcy)                                                                                                        \\ \hline
\end{tabular}
}
\vspace{-10pt}
\end{table}

\subsection{Multi-Path, Multi-Layer Exploration}
\begin{figure}
    \centering
    \includegraphics[width=0.85\columnwidth]{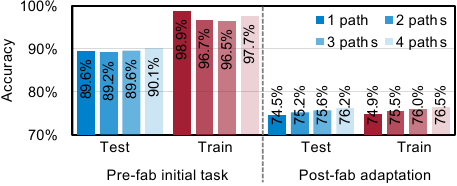}
    \vspace{-6pt}
    \caption{Adapt four DONNs with $L$=4 and different paths $P$ from pre-fab FMNIST task to post-fab Quickdraw task.
    Only permutation $\calP$ is trained during adaptation to show the learnability gain from enlarged solution space from 4! to 16!.}
    \label{fig:MetasurfacePaths}
\end{figure}

\begin{figure}
    \centering
    \includegraphics[width=0.75\columnwidth]{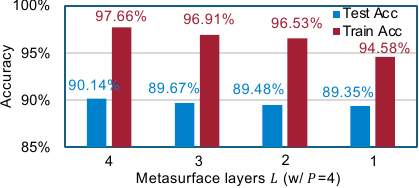}
    \vspace{-5pt}
    \caption{Different layers $L$ with $P$=4 on pre-fab FMNIST.}
     \vspace{-10pt}
    \label{fig:MetasurfaceLayers}
\end{figure}

We explore a more generic case with multiple parallel paths, enabling a larger design space and more functionalities.

\subsubsection{Number of Paths $P$ in Pre-Fab Initial Training}
In Fig.~\ref{fig:MetasurfacePaths}, we fix the layer count to $L$=4 and increase the number of diffractive paths $P$.
We observe approximately a 0.5\% improvement in test accuracy on the initial task. 
From the fluctuation in the training accuracy, we conclude that increasing too many parallel phase masks (similar to large-kernel convolution) can cause learning difficulty.

\subsubsection{Number of Paths $P$ in Post-Fab Task Adaptation with Permutation}
\label{subsec:PermutationAblation}
In Fig~\ref{fig:MetasurfacePaths}(\emph{Right}), we show task adaptation only with permutation learning.
By increasing the path from 1 to 4, the space of placement order increases from 4! to 16!, leading to a notable learnability boost and thus +1.7\% improvement on both test and training accuracy.

\subsubsection{Number of Layers $L$ in Pre-Fab Initial Training}
To investigate the impacts of metasurface layers on learnability, we conduct pre-fab training with 4 paths with different numbers of layers $L$. 
Figure~\ref{fig:MetasurfaceLayers} shows a clear benefit on expressivity with more layers cascaded. 

\subsection{Main Results}
We conduct three pre-fab training tasks on Fashion-MNIST, CIFAR-10, and Darcy, and transfer the \MS system to new tasks, i.e., QuickDraw, CIFAR-100, and Navier-Stokes, respectively. 
Table~\ref{tab:mainresults} demonstrates that \MS exhibits competitive expressivity compared to the digital counterpart in FMNIST and QuickDraw. 
However, we observe accuracy gaps on CIFAR-10 and CIFAR-100 due to the increased difficulty of those tasks.
However, we want to highlight that \MS \textbf{performs very well on much harder PDE prediction tasks on Darcy and Navier-Stokes, as those tasks require a global receptive field, which matches perfectly with the global spatial processing capability of diffractive ONNs}. 

For task adaptation, when all system variables \ding{202}-\ding{209} are enabled to learn, we can largely resume the accuracy even when the major weights $\Phi$ are frozen. 
To understand the individual contribution of each variable to task adaptability, we show a detailed discussion below.

\vspace{-5pt}
\subsection{Discussion}
\subsubsection{Parameter Contribution to Adaptability}
\begin{figure}
    \centering
    \vspace{-5pt}
    \includegraphics[width=0.85\columnwidth]{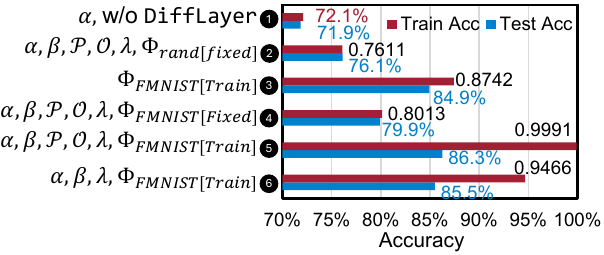}
    \vspace{-7pt}
    \caption{Training different parameters during adaptation.}
    \label{fig:PhaseAblationStudy}
     \vspace{-10pt}
\end{figure}

Figure~\ref{fig:PhaseAblationStudy} shows the insights of parameter contribution to adaptability. 

\noindent\underline{Compare (\ding{202}, \ding{205})}: Besides $\alpha_{pre}$, $\alpha_{post}$, \texttt{DiffLayers} are critical in DONN.

\noindent\underline{Compare (\ding{203}, \ding{205})}:
The trained $\Phi$ from initial tasks carries transferable knowledge that helps new task learning. 

\noindent\underline{Compare (\ding{205}, \ding{206})}:
If in the future, with advanced \emph{reconfigurable metasurfaces}, when $\Phi$ is trainable on new tasks even post-fabrication,
it indeed holds the great potential to enhance adaptability and learnability significantly. 

\noindent\underline{Compare (\ding{204}, \ding{206})}:
We want to highlight that even if $\Phi$ becomes trainable in the future, our introduced $\alpha,\beta,\calP,\calO, \lambda$ can still enhance the network's expressivity and learnability.

\noindent\underline{Compare (\ding{206}, \ding{207})}: 
It validates that our introduced permutation $\calP$ and orientation $\calO$ effectively expand the solution space and are crucial for enhancing the model's adaptability.

\subsubsection{Training Efficiency of Our Depth-wise \texttt{DiffLayer}} 
Compared to prior DONNs with 64-channel Conv-like diffractive layers, our depthwise \texttt{DiffLayer} consumes 29\% less memory (9807 MB to 6997 MB) and is 4.92$\times$ faster per training iteration (329.59 ms to 66.99 ms).

\subsubsection{Energy and Speed Advantages of \MS vs. Digital CNNs}
Via behavior-level architecture simulation, on average, \MS delivers 74$\times$ faster single-batch inference speed (1.48ms to 0.02ms) and consumes 194$\times$ less energy (134.9mJ to 0.696mJ) compared to a digital CNN on A6000 GPU. 
For the DONN part, we assume a photodetection (PD) speed of 10 GHz~\cite{NN_Optical23_Anderson}. 
Given that $P$ channels are processed in parallel, this results in an 8E-7 $ms$ latency with 32/$P$=8 ($P$=4) cycles and 0.1 ns per cycle. 
Thus, the latency of the diffractive layer is negligible compared to the overall time.

\begin{table}[H]
\centering
\vspace{-5pt}
\caption{Device info in energy and latency calculation~\cite{NP_ICCAD24_Gu}}
\vspace{-5pt}
\resizebox{0.9\columnwidth}{!}{
\begin{tabular}{c|c|c|c|c|c|c}
\hline
Device & 8-bit ADC & 8-bit DAC & TIA & MZM & PS & PD                                                 \\ \hline
Power (mW) & 14.8 & 50 & 3 & 50 & 0.15 & 0                                                 \\ \hline
\end{tabular}
\vspace{-15pt}
\label{tab:DeviceInfo}
}
\end{table}

For the PTC part, we employ a single-core 16$\times$16 PTC design~\cite{NP_ICCAD24_Gu}, with device power listed in Table~\ref{tab:DeviceInfo}. 
The energy and latency are simulated using an open-source photonic accelerator simulator~\cite{NP_HPCA2024_Zhu}.

\section{Conclusion}
\label{sec:conclusion}
We present the first in-depth analysis of multi-dimensional learnability in DONNs, introducing a physically composable hybrid optical/photonic system design, \MS. 
Our results demonstrate that \MS with multi-dimensional learnable system variables provides digital-comparable accuracy on various task adaptations, especially challenging PDE-solving tasks, with 74$\times$ faster inference speed and 194$\times$ lower energy.
Compared to prior DONNs, our hybrid \MS shows exponentially larger functional space and extreme multi-functionality with 5$\times$ faster training speed.
\MS shows the potential to enable versatile and high-efficiency optical AI systems with repurposed hardware, paving the way for practical deployment in dynamic AI applications.


\begin{thebibliography}{10}

\bibitem{NN_PhysRevLett19}
Tao Yan, Jiamin Wu, Tiankuang Zhou, Hao Xie, Feng Xu, Jingtao Fan, Lu~Fang, Xing Lin, and Qionghai Dai.
\newblock Fourier-space diffractive deep neural network.
\newblock {\em Phys. Rev. Lett.}, 123:023901, Jul 2019.

\bibitem{NP_ICCP24_Wei}
Kaixuan Wei, Xiao Li, Johannes Froech, Praneeth Chakravarthula, James Whitehead, Ethan Tseng, Arka Majumdar, and Felix Heide.
\newblock Spatially varying nanophotonic neural networks, 2023.

\bibitem{NP_OPTICA24_Peng}
Jingyang Peng, Li~Fang, Min Gu, and Qiming Zhang.
\newblock Dynamically reconfigurable all-optical neural network based on a hybrid graphene metasurface array.
\newblock {\em Opt. Continuum}, 3(5):704--713, May 2024.

\bibitem{NN_CHEN2021}
Hang Chen, Jianan Feng, Minwei Jiang, Yiqun Wang, Jie Lin, Jiubin Tan, and Peng Jin.
\newblock Diffractive deep neural networks at visible wavelengths.
\newblock {\em Engineering}, 7(10):1483--1491, 2021.

\bibitem{NN_Science17_Lin}
Xing Lin, Yair Rivenson, Nezih~T Yardimci, Mert Veli, Yi~Luo, Mona Jarrahi, and Aydogan Ozcan.
\newblock All-optical machine learning using diffractive deep neural networks.
\newblock {\em Science}, 361(6406):1004--1008, 2018.

\bibitem{NP_NATURE2017_Shen}
Yichen Shen, Nicholas~C. Harris, Scott Skirlo, et~al.
\newblock Deep learning with coherent nanophotonic circuits.
\newblock {\em Nature Photonics}, 2017.

\bibitem{NP_PIEEE2020_Cheng}
Q.~{Cheng}, J.~{Kwon}, M.~{Glick}, M.~{Bahadori}, L.~P. {Carloni}, and K.~{Bergman}.
\newblock {Silicon Photonics Codesign for Deep Learning}.
\newblock {\em Proceedings of the IEEE}, 2020.

\bibitem{NP_NaturePhotonics2021_Shastri}
Bhavin~J. Shastri, Alexander~N. Tait, et~al.
\newblock {Photonics for Artificial Intelligence and Neuromorphic Computing}.
\newblock {\em Nature Photonics}, 2021.

\bibitem{NP_SciRep2017_Tait}
Alexander~N. Tait, Thomas~Ferreira de~Lima, Ellen Zhou, et~al.
\newblock Neuromorphic photonic networks using silicon photonic weight banks.
\newblock {\em Sci. Rep.}, 2017.

\bibitem{NP_Nature2021_Xu}
Xingyuan Xu, Mengxi Tan, Bill Corcoran, Jiayang Wu, Andreas Boes, Thach~G. Nguyen, Sai~T. Chu, Brent~E. Little, Damien~G. Hicks, Roberto Morandotti, Arnan Mitchell, and David~J. Moss.
\newblock {11 TOPS photonic convolutional accelerator for optical neural networks}.
\newblock {\em Nature}, 2021.

\bibitem{NP_NatureComm2022_Zhu}
H.H. Zhu, J.~Zou, H.~Zhang, et~al.
\newblock Space-efficient optical computing with an integrated chip diffractive neural network.
\newblock {\em Nature Commun.}, 2022.

\bibitem{NP_ICCAD24_Gu}
Ziang Yin, Nicholas Gangi, Meng Zhang, Jeff Zhang, Rena Huang, and Jiaqi Gu.
\newblock Scatter: Algorithm-circuit co-sparse photonic accelerator with thermal-tolerant, power-efficient in-situ light redistribution.
\newblock In {\em International Conference on Computer-Aided Design (ICCAD)}, 2024.

\bibitem{NP_IJCAI23_Cunxi}
Yingjie Li, Weilu Gao, and Cunxi Yu.
\newblock Rubik's optical neural networks: multi-task learning with physics-aware rotation architecture.
\newblock In {\em Proceedings of the Thirty-Second International Joint Conference on Artificial Intelligence}, IJCAI '23, 2023.

\bibitem{NP_ICCAD22_Cunxi}
Yingjie Li, Ruiyang Chen, Weilu Gao, and Cunxi Yu.
\newblock Physics-aware differentiable discrete codesign for diffractive optical neural networks.
\newblock In {\em Proceedings of the 41st IEEE/ACM International Conference on Computer-Aided Design}, ICCAD '22, New York, NY, USA, 2022. Association for Computing Machinery.

\bibitem{NP_Science24_Lin}
Xing Lin, Yair Rivenson, Nezih~T. Yardimci, Muhammed Veli, Yi~Luo, Mona Jarrahi, and Aydogan Ozcan.
\newblock All-optical machine learning using diffractive deep neural networks.
\newblock {\em Science}, 361(6406):1004--1008, 2018.

\bibitem{goodman2005introduction}
Joseph~W Goodman.
\newblock {\em Introduction to Fourier Optics}.
\newblock Roberts and Company Publishers, 3rd edition, 2005.

\bibitem{NP_Nature2022_Luo}
Xuhao Luo, Yueqiang Hu, Xiangnian Ou, Xin Li, Jiajie Lai, Na~Liu, Xinbin Cheng, Anlian Pan, and Huigao Duan.
\newblock Metasurface-enabled on-chip multiplexed diffractive neural networks in the visible.
\newblock {\em Light: Science {\&} Applications}, 11(1):158, May 2022.

\bibitem{NP_Nature2024_Zheng}
Hanyu Zheng, Quan Liu, Ivan~I. Kravchenko, Xiaomeng Zhang, Yuankai Huo, and Jason~G. Valentine.
\newblock Multichannel meta-imagers for accelerating machine vision.
\newblock {\em Nature Nanotechnology}, 19(4):471--478, Apr 2024.

\bibitem{NP_Science2024_Xu}
Zhihao Xu, Tiankuang Zhou, Muzhou Ma, ChenChen Deng, Qionghai Dai, and Lu~Fang.
\newblock Large-scale photonic chiplet taichi empowers 160-tops/w artificial general intelligence.
\newblock {\em Science}, 384(6692):202--209, 2024.

\bibitem{NP_HPCA2024_Zhu}
Hanqing Zhu, Jiaqi Gu, Hanrui Wang, Zixuan Jiang, Zhekai Zhang, Rongxin Tang, Chenghao Feng, Song Han, et~al.
\newblock Lightening-transformer: A dynamically-operated photonic tensor core for energy-efficient transformer accelerator.
\newblock In {\em Proc.~HPCA}, 2024.

\bibitem{NP_ACS2022_Feng}
Chenghao Feng, Jiaqi Gu, Hanqing Zhu, Zhoufeng Ying, Zheng Zhao, et~al.
\newblock A compact butterfly-style silicon photonic--electronic neural chip for hardware-efficient deep learning.
\newblock {\em ACS Photonics}, 9(12):3906--3916, 2022.

\bibitem{NP_Nature2021_Feldmann}
Johannes Feldmann, Nathan Youngblood, Maxim Karpov, Helge Gehring, Xuan Li, Maik Stappers, Manuel~Le Gallo, Xin Fu, Anton Lukashchuk, Arslan Raja, Junqiu Liu, David Wright, Abu Sebastian, Tobias Kippenberg, Wolfram Pernice, and Harish Bhaskaran.
\newblock Parallel convolutional processing using an integrated photonic tensor core.
\newblock {\em Nature}, 2021.

\bibitem{NP_DAC2022_Gu}
Jiaqi Gu, Hanqing Zhu, Chenghao Feng, Zixuan Jiang, Mingjie Liu, Shuhan Zhang, Ray~T. Chen, and David~Z. Pan.
\newblock Adept: Automatic differentiable design of photonic tensor cores.
\newblock In {\em Proc.~DAC}, 2022.

\bibitem{NN_FashionMNIST2017}
Han Xiao, Kashif Rasul, and Roland Vollgraf.
\newblock {Fashion-MNIST: a Novel Image Dataset for Benchmarking Machine Learning Algorithms}.
\newblock {\em Arxiv}, 2017.

\bibitem{NN_QuickDraw2017}
Google.
\newblock The quick, draw! dataset, 2017.

\bibitem{NN_cifar2009}
Alex Krizhevsky, Geoffrey Hinton, et~al.
\newblock Learning multiple layers of features from tiny images.
\newblock 2009.

\bibitem{NN_DarcyFlow}
Zongyi Li, Nikola Kovachki, Kamyar Azizzadenesheli, Burigede Liu, Kaushik Bhattacharya, Andrew Stuart, and Anima Anandkumar.
\newblock Fourier neural operator for parametric partial differential equations, 2020.

\bibitem{NN_NaviorStoke}
Hyung-Chun~Lee John~Burkardt, Max~Gunzburger.
\newblock 2d navier stokes flow datasets.
\newblock 2006.

\bibitem{NN_AdamW2017_Ilya}
Ilya Loshchilov and Frank Hutter.
\newblock Decoupled weight decay regularization.
\newblock In {\em International Conference on Learning Representations}, 2018.

\bibitem{NN_ICLR2020_Esser}
Steven~K. Esser, Jeffrey~L. McKinstry, Deepika Bablani, Rathinakumar Appuswamy, and Dharmendra~S. Modha.
\newblock Learned step size quantization.
\newblock In {\em Proc.~ICLR}, 2020.

\bibitem{NN_Optical23_Anderson}
Maxwell~G. Anderson, Shi-Yuan Ma, Tianyu Wang, Logan~G. Wright, and Peter~L. McMahon.
\newblock Optical transformers, 2023.

\end{thebibliography}

\end{document}